\documentclass{article}

\usepackage{arxiv}
\usepackage{xcolor} 
\usepackage{colortbl} 
\usepackage{subcaption}
\usepackage{algorithm}
\usepackage{algorithmic}
\usepackage[utf8]{inputenc} 
\usepackage[T1]{fontenc}    
\usepackage{url}            
\usepackage{booktabs}       
\usepackage{amsfonts}       
\usepackage{nicefrac}       
\usepackage{microtype}      
\usepackage{amsmath}
\usepackage{lipsum}         
\usepackage{graphicx}
\usepackage{doi}

\title{Optimal Traffic Allocation for Multi-Slot Sponsored Search: Balance of Efficiency and Fairness}

\newif\ifuniqueAffiliation
\uniqueAffiliationtrue

\ifuniqueAffiliation 
\author{\hspace{1mm}Anastasiia Soboleva \\
	Avito \\ MSU Institute for Artificial Intelligence 
 \\ Moscow Institute of Physics and Technology\\
	\texttt{soboleva.an@phystech.edu} \\
	\And
	\hspace{1mm} Alexander Ledovsky \\
	Avito\\
	\texttt{adledovskiy@avito.ru} \\
	\AND
	  Yuriy Dorn\\
	MSU Institute for Artificial Intelligence \\ 
        Moscow Institute of Physics and Technology \\
	 \texttt{dornyv@my.msu.ru} \\
     \And
	  Egor Samosvat \\
	  Avito\\
	 \texttt{easamosvat@avito.ru} \\
     \AND
	  Andrey Tikhanov  \\
	  Moscow Institute of Physics and Technology\\
	 \texttt{tikhanov.ar@phystech.edu} \\
     \And
	  Fyodor Prazdnikov  \\
	  Avito\\
	 \texttt{fiprazdnikov@avito.ru} \\
}
\else
\usepackage{authblk}

\newbox{\orcid}\sbox{\orcid}{
\author[1]{%
	\href{https://orcid.org/0000-0000-0000-0000}{\usebox{\orcid}\hspace{1mm}David S.~Hippocampus\thanks{\texttt{hippo@cs.cranberry-lemon.edu}}}%
}
\author[1,2]{%
	\href{https://orcid.org/0000-0000-0000-0000}{\usebox{\orcid}\hspace{1mm}Elias D.~Striatum\thanks{\texttt{stariate@ee.mount-sheikh.edu}}}%
}
\affil[1]{Department of Computer Science, Cranberry-Lemon University, Pittsburgh, PA 15213}
\affil[2]{Department of Electrical Engineering, Mount-Sheikh University, Santa Narimana, Levand}
\fi

\renewcommand{\shorttitle}{\textit{arXiv} Template}

\hypersetup{
pdftitle={Optimal Traffic Allocation for Multi-Slot Sponsored Search: Balance of Efficiency and Fairness},
pdfauthor={Anastasiia Soboleva, Alexander Ledovsky},
}

\begin{document}
\maketitle
\begin{abstract}
The majority of online marketplaces offer promotion programs to sellers to acquire additional customers for their products. These programs typically allow sellers to allocate advertising budgets to promote their products, with higher budgets generally correlating to improve ad performance. Auction mechanisms with budget pacing are commonly employed to implement such ad systems. While auctions deliver satisfactory average effectiveness, ad performance under allocated budgets can be unfair in practice.

To address this issue, we propose a novel ad allocation model that departs from traditional auction mechanics. Our approach focuses on solving a global optimization problem that balances traffic allocation while considering platform efficiency and fairness constraints.

This study presents the following contributions. First, we introduce a fairness metric based on the Gini index. Second, we formulate the optimization problem incorporating efficiency and fairness objectives. Third, we offer an online algorithm to solve this optimization problem. Finally, we demonstrate that our approach achieves superior fairness compared to baseline auction-based algorithms without sacrificing efficiency. We contend that our proposed method can be effectively applied in real-time ad allocation scenarios and as an offline benchmark for evaluating the fairness-efficiency trade-off of existing auction-based systems.
\end{abstract}

\section{Introduction}
The growth of online marketplaces attracts both consumers seeking a vast array of products and sellers eager to capitalize on the increasing demand. To enhance product visibility, most marketplaces offer a variety of advertising options for sellers, including sponsored search listings and promotional placements within recommendation feeds.

The sponsored search involves a multi-step process: selection of candidate ads and choosing which ones to display. Major marketplaces commonly employ auction-based systems to handle ads allocation. Bidders can set bids manually or automatically through algorithms. Sellers often specify a budget, and pacing algorithms help distribute spending evenly over time.

Despite the widespread use and proven effectiveness of auction systems, traffic distribution among sellers can be uneven. This inequity arises primarily from two factors. Firstly, ad platforms prioritize products with high relevance and click-through rates (CTR) to maximize revenue. Secondly, real-world bid values often deviates from theoretical auction guarantees. Consequently, a small group of top-performing sellers disproportionately captures traffic at relatively low costs, while the majority of sellers struggle to gain significant visibility.

In this study we introduce Optimal Traffic Allocation (OTA) algorithm, a novel approach to ad allocation and ranking. OTA aims to balance fairness for advertisers in terms of budget utilization with platform efficiency in terms of the overall number of paid clicks. It offers a viable alternative to traditional auction-based systems with budget pacing.

Key features of this method include:
\begin{itemize}
    \item  Solves a convex optimization problem to balance fairness and efficiency
    \item Does not rely on auctions
    \item Assumes full budget spent, regardless of ad campaign outcome
    \item Operates independently of traffic patterns, ensuring consistent performance under diverse conditions
    \item Requires no historical data storage, reducing computational overhead
    \item Designed for efficient online implementation in high-traffic environments
\end{itemize}

Our mechanism operates as follows. When a user submits a query, the OTA algorithm receives relevant items, their estimated click-through rates, allocated budgets, and available ad slots. By solving a convex optimization problem, the algorithm determines the optimal distribution of target actions among these items. Finally,  probabilistic ranking assigns candidate ads to the available advertising slots.

OTA is well-suited for real-time advertising systems. Additionally, it can serve as an offline benchmark to evaluate the fairness-efficiency trade-off of existing auction-based platforms. It is worth noting that many existing approaches require additional input parameters beyond budget, such as maximum click price. While control over click prices offers advertisers a degree of flexibility, it may be overly complex for average marketplace sellers who are unfamiliar with their target click price. For such sellers, a more straightforward approach might involve only two controls: the daily budget and the campaign length, where the daily budget serves as a kind of bid. Through our work on promoting products on *** marketplace, we've found this design to be highly effective. Therefore, the design of OTA algorithm can be particularly beneficial for small marketplace sellers.

The rest of the paper describes specific aspects of the OTA algorithm. This includes formulating the optimization problem with a Gini-index based fairness metric, developing an efficient O(N) algorithm based on the Frank-Wolfe method, and conducting comprehensive experiments comparing OTA to baseline methods. Our results demonstrate OTA's superior fairness without sacrificing efficiency.
\section{Related work}
We explore two primary areas of research relevant to our work: comparable advertising mechanisms and fairness objectives in advertising.

\textbf{Comparable advertising mechanisms.} In this study, we propose a budget-based advertising approach that ensures full budget utilization. In marketplaces, auction-based advertising systems typically employ budget pacing to manage budgets \cite{Chen2024, Nguyen2023PracticalBP}. Budget pacing is a class of solutions designed to satisfy daily budget goals and ensure smooth spending throughout the time period.

Although various pacing algorithms exist, they differ in their working principles. Hard pacing (or throttling) \cite{Agarwal2014, Karande2013, Nguyen2023PracticalBP} controls the probability of participating in auctions and requires bid values as input. Bid modification (or soft pacing) in most realizations expects bids or target action values as well (\cite{Stram2024, Chen2024}). Some implementations also incorporate return-on-investment (ROI) constraints \cite{Lucier2023AutobiddersWB}.

Beyond traditional auctions, our research is closely tied to the field of online allocation. A comprehensive review of this area can be found in \cite{balseiro2020}. Online allocation presents an optimization problem to be solved in real-time upon new search requests, with a focus on maximizing reward functions under budget constraints. The primary distinction between our work and online allocation lies in the need for a predefined resource consumption function in online allocation. This requires dynamic bids (essentially creating an auction) or fixed target action values. In contrast, our problem formulation does not require such a function but instead exploiting fairness objectives to balance traffic between ads.

An advertising product design that shares similarities with our approach is Guaranteed Display (GD) advertising (\cite{Yang2010InventoryAF}). In GD, advertisers pay a fixed budget for a specified number of target actions (typically impressions). If the advertiser's goal is not met, they typically receive a refund. Some pacing approaches have been developed for GD, such as \cite{dai2024}, which uses Dual Mirror Descent (\cite{balseiro2020}).

However, GD differs from our problem formulation in that impression goals serve as the budget constraints. This means that fairness is managed automatically when signing fair contracts with advertisers. However, ensuring good platform efficiency in terms of overall paid traffic becomes more challenging due to idle sponsored spots when advertiser goals are satisfied and lack of consideration for ad quality (CTRs).

\textbf{Fairness objectives} in optimization problems have been defined and implemented in various ways. A critical survey of different schemes for formulating fairness criteria in optimization models can be found in \cite{XinyingChen2023}. One widely used approach for combining efficiency and fairness is the $\alpha$-fairness scheme, introduced in \cite{Atkinson1970}. This scheme generalizes some special cases of fairness, including max-min and proportional fairness. The trade-off between fairness and efficiency based on $\alpha$-fairness was studied in \cite{Bertsimas2012}, where theoretical estimates of the price of fairness were obtained. \cite{Bateni2022} proposed an approach that employs proportional fairness to allocate resources in scenarios with Gaussian-distributed input traffic.

However, it is worth noting that the $\alpha$-fairness scheme defines both efficiency and fairness using a single utility function, which may be a limitation in practice. In contrast, our work defines efficiency as the overall number of paid clicks and fairness as a function of impressions per unit budget. Other approaches, such as regularized online allocation problem introduced in \cite{Balseiro2021}, can support various fairness constraints. This approach is based on relative consumption and requires a consumption function.

The Gini Mean Difference (GMD) is a well-known measure of inequality, first proposed by \cite{gini1912}. The GMD has been extensively applied in various fields, with numerous practical applications and modifications reported in the literature \cite{Yitzhaki2012}. Geometrically, the GMD can be interpreted as proportional to the area between the Lorenz curve and a diagonal line representing perfect equality. GMD has also been applied to advertising-related problems, such as ad inventory allocation. \cite{Lejeune2019} presented an optimization model with a Gini-based objective function to allocate ad inventories efficiently. Their work demonstrated the effectiveness of using GMD in this context and provided an efficient algorithm for solving the resulting optimization problem. Notably, ad inventory allocation differs from online ad allocation in several key aspects: it operates offline, whereas online ad allocation occurs in real-time; and it focuses on audience segments rather than individual users or search queries. Besides its applications in advertising, the GMD has also been applied in ranking problems, particularly in \cite{do2021two, do2022optimizing}.

Other notable approaches to ensuring fairness in advertisement systems include the two-stage algorithm for online allocation introduced in \cite{Li2024}. This approach involves an offline stage where the desired amount of clicks on each item from each user segment is estimated, followed by an online optimization problem that chooses an optimal policy to display ads.

\section{Problem Formulation}

We consider a sponsored search allocation problem with finite horizon of  $T$ incoming search requests. For each request $t\in [T]$, the platform provides $K(t)$ potential advertising slots. We consider a set of $N$ items, each associated with a non-negative budget $B_j > 0$, where $j = 1, \ldots, N$. These items can be placed in available slots. We denote $d_{jt}^k$ as the binary indicator of displaying item $j$ on the slot $k$ for a search request $t \in T$. We define the probability of click on displaying item as $\gamma_k \cdot c_{jt}$ where  $c_{jt}$ is Click-Through Rate (CTR) and  $\gamma_k$ is a position multiplier \cite{chuklin2022click}. CTR $c_{jt} \in [0, 1]$ is defined as probability of click if the item $j$ was observed. $\gamma_k \in [0,1]$ is defined as probability that user observes the ad at the $k$-th position and $1 \geq \gamma_1 > \gamma_2 > ... > \gamma_{K(t)} \geq 0$. We expect that CTR and position multipliers are known to the platform. For convenience, we refer to the expression $\gamma_k d^k_{jt}$ as an "impression".

\textbf{Utility and Fairness.}  We define the utility $U_j(d)$ of the slot's allocation $d = \{d_{jt}^k |  j\in [N], t \in [T], k \in [K(t)]\}\ $ for the advertiser $j \in [N]$ as the total number of impressions they receive, given by:
$$U_j(d) = \sum_{t\in [T]}\sum_{k\in [K(t)]}\gamma_k d^k_{jt} \eqno(1)$$

To measure fairness, we propose using of Gini Mean Difference (GMD), defined as:
$$GMD(d) = \frac{1}{N^2}\sum_{j,h\in [N]}\left| \frac{U_j(d)}{B_j}-\frac{U_h(d)}{B_h}\right| = $$
$$ = \frac{1}{N^2}\sum_{j,h\in [N]} \left| \sum_{t\in [T]}\sum_{k\in [K(t)]}\gamma_k \left(\frac{d^k_{jt}}{B_j}-\frac{d^k_{ht}}{B_h}\right)\right| \eqno(2)$$

The GMD is proportional to the Gini index, defined as:
$$ Gini(d)  = \frac{GMD(d)}{2\mu}, \eqno(3)$$ 
where $\mu =\frac{1}{N} \sum_{j\in [N]} \frac{U_j(d)}{B_j}$ is the average impression cost among all items. Gini index ranges from 0 (perfect equality) to 1 (absolute inequality).

\textbf{Efficiency.}
We define efficiency as the total number of clicks on promoted items, which can be calculated as:

$$E(d) = \sum_{t\in [T]}\sum_{k\in [K(t)]}\sum_{j\in [N]} c_{jt} \gamma_k  d^k_{jt} \eqno (5) $$
This formulation can be easily extended to other types of target actions by replacing CTR model with a corresponding conversion model.

\textbf{Optimization problem.} We propose an optimization problem that considers both efficiency and fairness components:
$$\max_{d} \hspace{5pt}(1-\lambda) \cdot E(d) - \lambda \cdot GMD(d) \eqno(6)$$

 s.t $$\sum_{j\in[N]} d_{jt}^k \leq 1 \hspace{7pt} \forall t \in [T],\forall k \in [K(t)] \eqno(6.a)$$
 
$$\sum_{k\in[K]} d_{jt}^k \leq 1 \hspace{7pt}  \forall t\in[T],\forall j\in[N]\eqno(6.b)$$

$$d_{j t}^k \in \{0,1\} \hspace{7pt} \forall t\in[T],\forall k\in[K(t)], \forall j\in[N]$$

Where $\lambda \in [0, 1]$ is a parameter that defines a trade-off between efficiency and fairness. $(6.a)$ ensures that each slot can have only one item. $(6.b)$ ensures that item can be displayed only in one slot at the same time. 

Unfortunately, the problem $(6)$ has several practical limitations. First, it requires prior knowledge of the per-query traffic data which is unachievable in reality. Second, the proposed binary optimization program appears to be computationally hard, making it challenging to solve efficiently.

Despite the difficulties described above, the following sections present a revised approach that mitigates these issues. 

\section{Reformulation of the Problem}

To address the challenges associated with the original optimization problem, we propose a two-stage model for Optimal Traffic Allocation (OTA). The first stage involves an algorithm that finds the optimal theoretical distribution of impressions $\Gamma(t)$ across items on a specific search request $t\in [T]$. At the second stage, a probabilistic ranking approach is employed to allocate final slots to items based on their predicted performance probabilities. Both stages can be executed in real-time for each incoming search query.

Drawing inspiration from the idea presented in \cite{Li2024} on transforming the allocation problem into a distribution problem, we apply a similar approach. We define the total amount of impressions on a specific search request as $\Gamma(t) = \sum_{k=1}^{K(t)} \gamma_k$. We consider the scenario $(N\geq K(t))$ where the number of items exceeds the number of available slots. For a given search query $t$, we introduce a virtual impression vector:

$$\alpha := (\alpha_{j}, j\in[N]) \in \mathbb{R}^{N}$$
$$0\leq\alpha_{j}\leq 1$$$$\sum_{j \in [N]} \alpha_{j} = \Gamma(t)$$

The virtual impression vector abstracts away from the fact that real impression values are constrained by possible slot allocations and position multipliers, while still respecting other relevant constraints.

For a particular search query $t\in [T]$ the GMD $(2)$ and Gini index $(3)$  are converted as follows:
$$GMD_q(\alpha) = \frac{1}{N^2} \sum_{j,h\in [N]}\left| \frac{\alpha_{j}}{B_j}-\frac{\alpha_{h}}{B_{h}}\right| \eqno(8)$$
$$Gini_q(\alpha) = \frac{GMD_q(\alpha)}{2\mu} \eqno(9) $$
where $\mu =\frac{1}{N} \sum_{j\in [N]} \frac{\alpha_j}{B_j}$ is the average virtual impression cost among all items. 

Optimization of $GMD_q$ (8) still can be challenging, as it involves solving the problem of minimizing modulus. Because of that we propose $G_q(\alpha)$ $(10)$. This objective exhibits similar behavior to the original GMD metric but with significantly improved computational performance.
$$G_q(\alpha) = \frac{1}{N^2} \sum_{j,h\in [N]}\left( \frac{\alpha_{j}}{B_j}-\frac{\alpha_{h}}{B_h}\right)^2 \eqno(10)$$
In addition to our fairness objective, we also introduce a per-query efficiency objective:
$$E_q(\alpha) = \sum_{j\in [N]} c_{j} \cdot \alpha_{j}  \eqno(11)$$

\textbf{First Stage Optimal Traffic Distribution Optimization Problem.} Combining new fairness and effectiveness objectives with virtual impression vector we introduce a new per-query optimization problem:
$$\max_{\alpha}\hspace{5pt} (1 - \lambda)\cdot E_q(\alpha) - \lambda \cdot G_q(\alpha) \eqno(12)$$
s.t 
$$\sum_{j\in [N]}\alpha_{j} = \sum_{k}\gamma^k \hspace{5pt}$$
$$\alpha_j \leq 1 \hspace{5pt} \forall j\in [N]$$
$$\alpha_j \geq 0 \hspace{5pt} \forall j\in [N]$$

Where $\lambda \geq 0 $ defines the trade-off between efficiency and fairness identically to $(6)$.

The problem $(12)$ can be reformulated as a quadratic programming problem in standard form through the transformations outlined in appendix:
$$\min_{0\leq x\leq 1} \frac{1}{2}\lambda\left<Bx,x\right> - (1-\lambda)\left<c,x\right> \eqno(13)$$
s.t$$ \left<x,e\right> = \Gamma\eqno(14)$$
where $e\in \mathbb{R}^N$ is a unit vector, $x\in \mathbb{R}^N$, $B_{N\times N}$ is a symmetric positive definite matrix dependent only on budgets, $c\in \mathbb{R}^N = (c_{1}, c_{2},...,c_{N})$ represents a vector of CTR values, $N$ denotes the number of items participating in the request, $\Gamma = \sum_{k\in [K(t)]}\gamma_k$.

To explain the transition from a quadratic programming problem $(12)$ to one in matrix form $((13)-(14))$, we rewrite the expression as follows:

$$\frac{1}{N^2}\sum\limits_{j,h\in [N]}\left( \frac{x_j}{B_j}-\frac{x_h}{B_h}\right)^2 = 
\frac{1}{N^2}\sum\limits_{j,h\in [N]}\left( \frac{x^2_j}{B^2_j}+\frac{x^2_h}{B^2_h} - 2\frac{x_j}{B_j} \frac{x_h}{B_h}  \right)=$$
$$=\frac{1}{N^2}\left(2N\sum\limits_{j\in [N]}\frac{x^2_j}{B^2_j} - \sum\limits_{j,h\in [N]}2\frac{x_j}{B_j} \frac{x_h}{B_h}\right)  = \frac{1}{2}\left<Bx,x\right>,$$
where matrix $B$ defined as: 

\begin{equation*}
B_{jh} = 
 \begin{cases}
   -\frac{4}{N^2}\frac{1}{B_j B_h} &\text{if $j \neq h$}\\
    \frac{4(N-1)}{N^2}\frac{1}{B_j^2} &\text{if $j = h$}
 \end{cases}
\end{equation*}

\section{Algorithms}
As mentioned before, the OTA model consists of two stages. In the first stage, we focus on solving the optimization problem $(12)$ to obtain an optimal impression distribution. Subsequently, in the second stage, we employ a probabilistic ranking approach that aims to approximate this optimal distribution in expectation.

\subsection{Optimal impression distribution stage.} The optimization problem $(12)$, transformed to the form $(13)-(14)$, is a quadratic programming problem with constraints. Several well-established methods exist for solving such problems, which can be broadly classified into two categories: gradient descent methods with projection and the Frank-Wolfe method \cite{bubeck2015convex}. From a practical standpoint, both methods can yield satisfactory results. However, we advocate using the Frank-Wolfe method ($Algorithm~1$) due to its inherent suitability for our problem.

The Frank-Wolfe method's effectiveness relies on being able to efficiently solve an internal optimization problem, specifically $\min\limits_{s \in D} \langle f'(x), s \rangle$ which is embedded in $Algorithm~1$ in  $line~2$ in the following form: $$\min_{s\in D} \langle \lambda Bx-(1-\lambda) c, s \rangle\eqno(15)$$
s.t
$$D = \{s\in \mathbb{R}^N: 0 \leq s \leq 1 \& \left<s, e\right>= \Gamma\}\eqno(16)$$
Solving problem $(15)-(16)$ in our case is straightforward, with a solution that can be computed directly: it consists of selecting the top $K$ components of the vector $\lambda Bx-(1-\lambda)c$.

\begin{algorithm}
	\caption{OTD-FW}
	\label{alg:algorithm}
	\renewcommand{\algorithmicrequire}{\textbf{Input:}}
	\renewcommand{\algorithmicensure}{\textbf{Output:}}
	\begin{algorithmic}[1]
		\REQUIRE Budget matrix $B_{N\times N}$, conversion vector $c\in \mathbb{R}^N$, maximum number of advertising slots $K$. \textbf{Parameters:} Trade-off parameter $0\leq\lambda \leq 1$, maximum number of iterations $T$, position multipliers $(\gamma_k)_{k\in[K]}$,  $\Gamma = \sum_{k=1}^K \gamma_k$, initial point $x_0 \in D$, $D = \{s\in \mathbb{R}^N \quad | \quad 0 \leq s \leq 1, \quad  \left<s, e\right> =  \Gamma\}.$
		\ENSURE Distribution vector $x \in \mathbb{R}^N$
		\FOR{$t = 0, \ldots, T$} 
        \STATE Compute $s_t = argmin_{s\in D}\left<\lambda Bx_{t-1} - (1-\lambda)c,s\right>$\\
        \STATE Set $\beta_t = \frac{2}{t+2}$\\
        \STATE Compute $x_t = (1-\beta_t) x_{t-1} + \beta_t s_t$\\
        \ENDFOR
        \STATE \textbf{return} $x_T$
	\end{algorithmic}
\end{algorithm}

\textbf{Theorem} (\cite{bubeck2015convex}). Let $f$ be a convex and $\sigma$-smooth function w.r.t. some norm $||\cdot||$, $R=\sup\limits_{x,y \in D} ||x-y||$, and $\beta_t  = \frac{2}{t+2}$ for $t \geq 0$.Then for any $t \geq 1$, one has $$f(x_t) - f(x^*)\leq \frac{2 \sigma R^2}{t+2.}$$
You can see the proof of the theorem in \cite{bubeck2015convex} ($Theorem~3.8$). 

\textbf{Proposition.} The $Theorem~3.8$ ($\cite{bubeck2015convex}$) implies the following convergence of $Algorithm~1$ in relation to our problem $(13)-(14)$:$$f(x_t) - f(x^*)\leq \frac{2 \lambda ||B||_{\infty}}{t+2}\hspace{10pt} \forall t \geq1 $$ where $f(x) = \frac{1}{2}\lambda\left<Bx,x\right> - (1-\lambda)\left<c,x\right>$, $x_t$ corresponds to $x$ on iteration $t$, $x^*$ is an optimal solution $(13)-(14)$, and $||B||_{\infty} = \max\limits_{1\leq i\leq N}\sum_{j=1}^N|b_{ij}|$. 

The proposition above allows to estimate the required number of iterations to achieve the desired accuracy. Note that in practice, the algorithm achieves the required accuracy in fewer iterations.

\textbf{Proof of Proposition.} Let's relate $Theorem~3.8$ ($\cite{bubeck2015convex}$) to our problem $(13)-(14)$. We focus on the maximum norm, denoted as $||\cdot||_{\infty}$. For a vector $x$, the maximum norm is given by: $$||x||_{\infty} := \max\limits_{1\leq k\leq n} |x_k|$$ and the subordinate matrix infinity norm is defined as: $$||A||_{\infty} = \max\limits_{1\leq i\leq m} \left( \sum\limits_{j=1}^n |a_{ij}| \right)$$

In our problem $(13) - (14)$
$$f(x) = \frac{1}{2}\lambda\left<Bx,x\right> - (1-\lambda)\left<c,x\right>$$
$$f'(x) = \lambda Bx - (1-\lambda)c$$
$$D = \{s\in \mathbb{R}^N: 0 \leq s \leq 1 \& \left<s, e\right> \leq \Gamma\}$$
$$R=\sup\limits_{x,y \in D} ||x-y||_{\infty} = \max\limits_{x,y \in D}(|x_1-y_1|,...,|x_n - y_n|) \leq 1$$
$$||f'(x) - f'(y)||_{\infty} = ||\lambda B(x-y)||_{\infty}\leq \lambda ||B||_{\infty}||x-y||_{\infty}, \forall x, y \in D$$
So $f(x)$ is a $||B||_{\infty}$ - smooth function. According to the $Theorem~3.8$ ($\cite{bubeck2015convex}$) $$f(x_t) - f(x^*)\leq \frac{2 \lambda ||B||_{\infty}}{t+2}\hspace{10pt} \forall t\geq 1$$ That completes the proof of our $Proposition$.

\subsection{Probabilistic ranking stage.}
\begin{algorithm}
\caption{OTA}
\label{alg:algorithm2}
\renewcommand{\algorithmicrequire}{\textbf{Input:}}
\renewcommand{\algorithmicensure}{\textbf{Output:}}
\begin{algorithmic}[1] 
        \REQUIRE Budget matrix $B_{N\times N}$, conversion vector $c\in \mathbb{R}^N$, maximum number of advertising slots $K$. \textbf{Parameters:} Trade-off parameter $0\leq\lambda \leq 1$, maximum number of iterations $T$, visibility coefficients $(\gamma_k)_{k\in[K]}$.
		\ENSURE The final allocation of slots between sellers.
        \STATE Find the optimal distribution $\alpha$ by $OTD-FW$ with input parameters
        \STATE Initialize set $L_1 = \{1,...,N\}$
        \FOR{$k = 1,...,K$}
        \STATE Sample index $l_k \in [L_k]$ from distribution \\$\mathbb{P}[d_l^k = 1] \sim \alpha[l], \quad \forall l \in [L_k]$ \\
        \STATE Display an item $l_k$ on slot $k$\\ 
        \STATE Update the set of items under consideration\\ $L_{k+1} = L_k\setminus \{l_k\}$
        \IF {$L_k = \emptyset$}
        \STATE Break
        \ENDIF
        \ENDFOR
\end{algorithmic}
\end{algorithm}  

During Stage 1, our optimal distribution algorithm operates in real-time for each search query, generating a desired distribution of impressions across all items. However, this distribution does not align with the available slot allocations. To address this discrepancy, we introduce the probabilistic ranking stage which employs randomization to allocate items to slots such that the desired distribution is approximated in expectation. To achieve this, we propose an algorithm which samples items for allocation to slots based on probabilities proportional to the distribution computed by the Frank-Wolfe method.

The full description of the two-stage Optimal Traffic Allocation algorithm is presented in $Algorithm~2$. This end-to-end process executes for every individual search query. Thus, OTA, an online algorithm, optimizes a global objective function based on the Gini index. While it employs budgets, it does not facilitate direct withdrawal transactions.

\section{Experimental Results}
We evaluate our algorithm using both synthetic and real-world data from *** marketplace production logs. Synthetic data experiments are designed to verify the convergence of OTA's probabilistic ranking to its OTD-FW's theoretical values, ensuring the algorithm's correctness. Real-world data experiments, on the other hand, aim to compare the efficiency and fairness trade-offs of OTA with alternative approaches, enabling us to validate its practical effectiveness.

\subsection{Experiments on Synthetic Data}
\begin{figure*}[h]
  \includegraphics[width=1\textwidth]{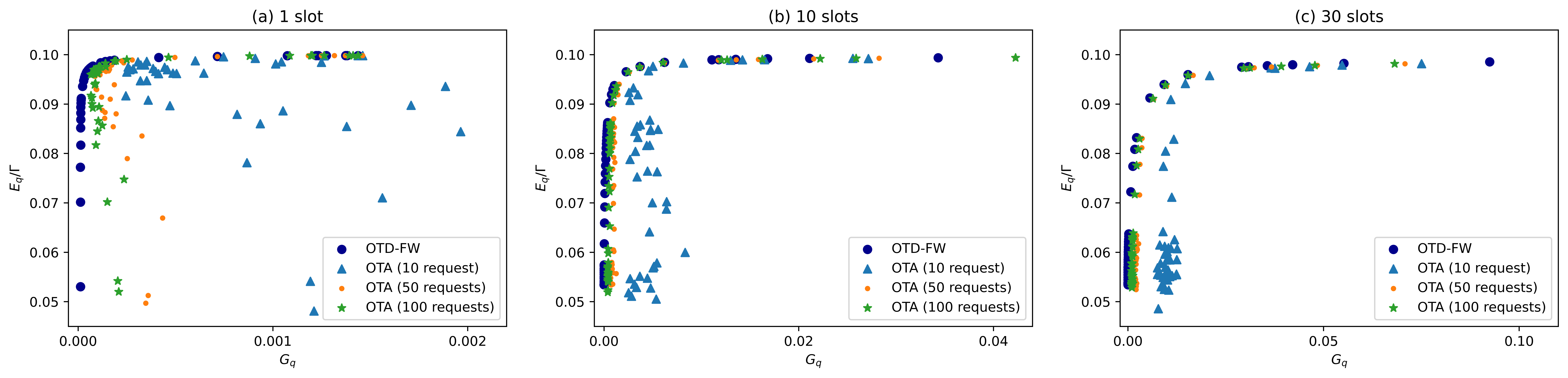}
  \caption{Convergence of $OTA$ to the theoretical distribution by $OTD-FW$}
  \label{fig:teaser}
\end{figure*}
To verify the correctness of OTA's probabilistic ranking we created a synthetic dataset consisting of a single repeated query extracted from our production logs. The number of items $N$ was set to $1000$. The number of repeated queries varied in three scenarios: $[10, 50, 100]$. Additionally we explored three distinct slot configurations: $[1, 10, 30]$.

We executed the $OTD-FW$ algorithm on the test query obtaining a set of virtual impression values $\alpha$. To assess the accuracy of our probabilistic ranking, we compared two key metrics: $E_q(\alpha) / \Gamma$ and $G_q(\alpha)$. We computed their theoretical values and these same metrics after applying the probabilistic ranking.

As depicted in $Figure~1$, our experiments demonstrate that with $100$ queries, we achieve a close approximation to the theoretical metrics. Furthermore, for scenarios involving $10$ and $30$ slots, even as few as $10$ queries are sufficient to produce accurate results.

\subsection{Experiments on Real Data}
We collected an industrial dataset based on *** marketplace production logs. The dataset consists of five day data in one product category in specific location resulting with $25K$ search queries and $200$ ad campaigns. Each query include a list of ad candidates with their CTRs and number of ad slots.

\subsubsection{Metrics} 
We employed fairness and efficiency metrics to evaluate the performance of our proposed OTA algorithm in comparison with baseline algorithms. As a metric of fairness, we evaluated the Gini index $(3)$, which quantifies the distribution of total impressions among items:
$$ Gini = \frac{\sum\limits_{j,h\in [N]} \left| \sum\limits_{t\in [T]}\sum\limits_{k\in [K]}\gamma_k \left(\frac{d^k_{jt}}{B_j}-\frac{d^k_{ht}}{B_h}\right)\right|}{2N\sum\limits_{j\in [N]} \sum\limits_{t\in [T]}\sum\limits_{k\in [K]}\gamma_k \frac{d^k_{jt}}{B_j}}$$
As a metric of efficiency we evaluated the expected number of clicks per query $(5)$:
$$Efficiency = \frac{1}{T}\sum\limits_{t\in [T]}\sum\limits_{k\in [K]}\sum\limits_{j\in [N]} c_{jt} \gamma_k  d^k_{jt}$$
In evaluation results we presented the efficiency as a relative value, normalized to the performance of CTR-ranking.

\subsubsection{Baselines}

\begin{figure}[ht]
    \centering
    \includegraphics[width=0.8\textwidth]{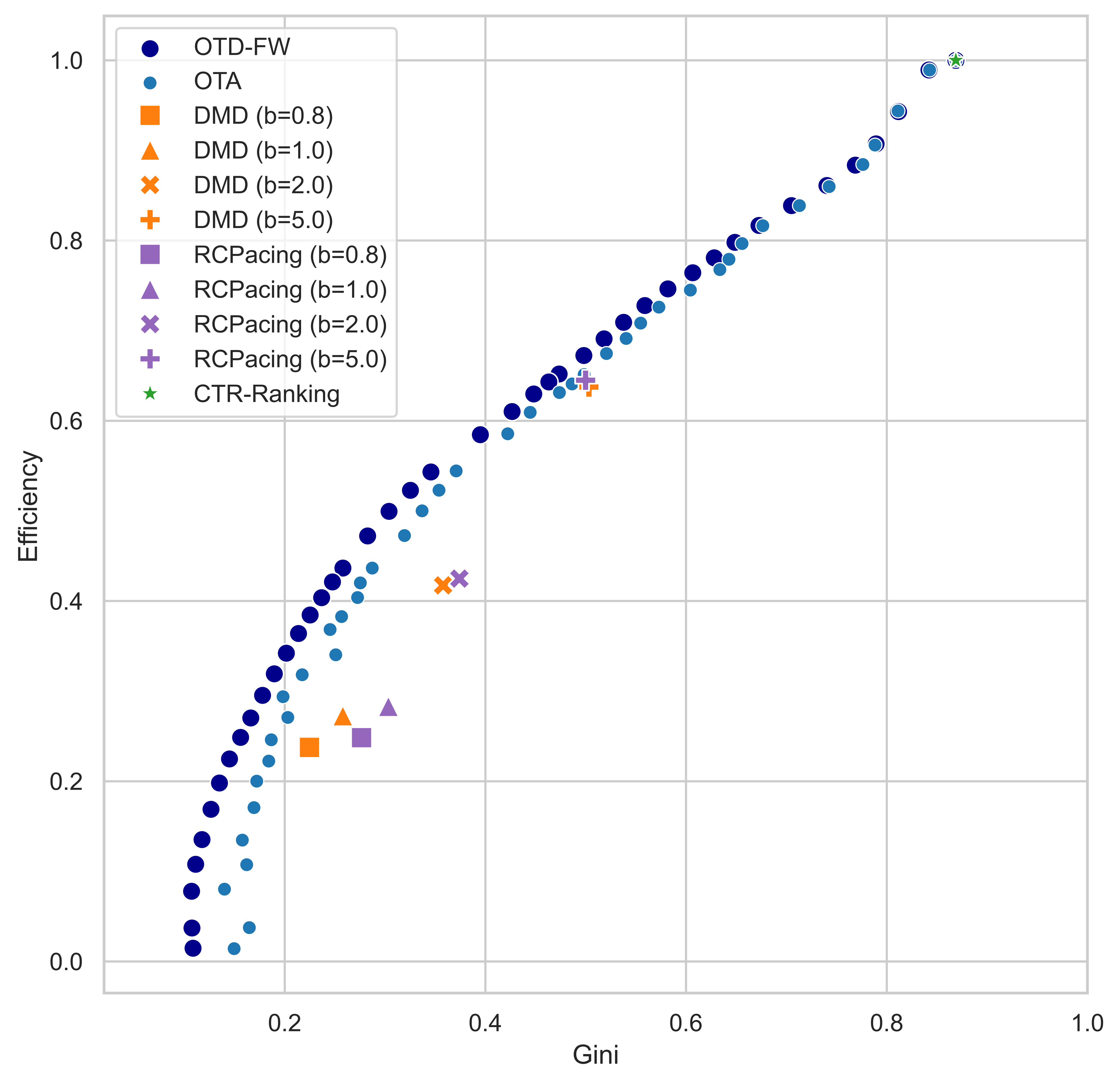}
    \caption{Values of Efficiency and Gini metrics for the following algorithms: OTA, RCPacing, DMD and CTR-ranking. OTA algorithm is provided for different $\lambda$. PRPacing and DMD are provided for different impression amounts within budgets. The efficiency metric is presented as a relative value, normalized to the performance of CTR-ranking.}\label{fig:animals}
\end{figure}

We compare \textbf{OTA} with three other methods.  \textbf{RCPacing} - a pacing algorithm for guaranteed display advertising, that utilizes Lagrangian dual multipliers to fine-tune probabilistic throttling through monotonic mapping function within the percentile space\cite{dai2024}. \textbf{DMD} - an online allocation framework that uses Lagrangian-based optimization to generate virtual bids for ad items \cite{balseiro2020}. We choose to compare with DMD and RCPacing because of the potentially good fairness in GD environments. \textbf{Ranking by CTR} - a baseline that achieves the maximum possible efficiency by design.

We obtained implementations of RCPacing and DMD algorithms from the supplementary github repository of \cite{dai2024}. As previously discussed, GD pacing interprets budget constraints as targets for the number of impressions. To achieve meaningful comparisons, we evaluated several scenarios with varying total impression amounts within GD budgets. Besides of that, because of RCPacing and DMD implementations, we conducted the simulation at $1$ ad slot design.

\subsubsection{Evaluation results} 

The evaluation results presented in $Figure~ 2$ demonstrate that OTA produced results that were close but slightly divergent from the theoretical values calculated using OTD-FW, due to approximation errors in the probabilistic ranking process. Nevertheless, OTA demonstrated a superior trade-off compared to RCPacing and DMD, with approximate Gini index reduction of \textbf{23.1\%} (DMD) and \textbf{34.6\%} (RCPacing) while maintaining comparable efficiency. Although the uplift may seem significant, it is deemed acceptable given OTA's primary objective of balancing fairness and effectiveness. Therefore, this experiment provides a strong evidence that OTA works as expected.

Additionally, it can be observed that there are areas where a good balance between efficiency and fairness can be achieved by adjusting the $\lambda$ parameter from $0.8$ to $0.9$ when the Gini index is still relatively low $(0.3-0.4)$ and efficiency is relatively high comparing to CTR-Ranking $(0.5-0.6)$. In these areas, the advertising platform may aim to maintain its positioning.

\begin{table}[h]
\centering
\begin{subtable}[h]{0.45\textwidth}
\begin{tabular}{|c|c|c|}
\hline
\rowcolor[gray]{0.9}
 & \multicolumn{2}{c|}{\textbf{$\Delta$ Gini}} \\
\hline
\rowcolor[gray]{0.9}
 & OTD-FW & OTA \\
\hline
DMD (b = 0.8) \hspace{11pt}  & -30.6\% & -16.9\% \\ \hline
DMD (b = 1)\hspace{20pt}     & -31.0\% & -23.1\%\\ \hline
DMD (b = 2)\hspace{20pt}     & -30.8\% & -23.0\% \\ \hline
DMD (b = 5)\hspace{20pt}     & -7.9\% & -3.3\%\\ \hline
\rowcolor[gray]{0.9}
& &  \\ \hline 
RCPacing (b=0.8)           & -35.6\% & -28.3\%\\ \hline 
RCPacing (b=1)\hspace{6pt} & -41.3\% & -34.6\%\\ \hline
RCPacing (b=2)\hspace{6pt} & -31.1\% & -23.2\%\\ \hline
RCPacing (b=5)\hspace{6pt} & -5.3\% & -0.3\%\\ \hline
\end{tabular}
\caption{Gini Improvements.}
\end{subtable}
\hfill
\begin{subtable}[h]{0.45\textwidth}
\centering
\begin{tabular}{|c|c|c|}
\hline
\rowcolor[gray]{0.9}
 & \multicolumn{2}{c|}{\textbf{$\Delta$ Efficiency}} \\
\hline
\rowcolor[gray]{0.9}
& OTD-FW & OTA \\
\hline
DMD (b = 0.8) \hspace{11pt}   & +34.4\% & +33.9\% \\ \hline
DMD (b = 1)\hspace{20pt}      & +41.1\% & +40.4\%\\ \hline
DMD (b = 2)\hspace{20pt}      & +25.3\% & +25.3\% \\ \hline
DMD (b = 5)\hspace{20pt}      & +2.3\% & +2.2\%\\ \hline
\rowcolor[gray]{0.9}
&  & \\ \hline 
RCPacing (b=0.8)            & +69.6\% & +69.2\%\\ \hline 
RCPacing (b=1)\hspace{6pt}  & +54.3\% & +54.2\%\\ \hline
RCPacing (b=2)\hspace{6pt}  & +27.9\% & +28.1\%\\ \hline
RCPacing (b=5)\hspace{6pt}  & +1.1\% & +1.0\%\\ \hline
\end{tabular}
\caption{Efficiency Improvements.}
\end{subtable}
\end{table}

In \textit{Table 1}, we present the comparative analysis of Gini indices between our proposed OTA algorithm and baseline methods under fixed efficiency. This assessment includes theoretical estimations from OTD-FW that serve as potential values for the OTA approach. In \textit{Table 2}, we present a similar evaluation for the efficiency metric under fixed Gini.

\section{Conclusion}

The proposed Optimal Traffic Allocation algorithm offers a novel ranking approach for ad slots, providing an alternative to traditional auction-based methods. Unlike auctions, OTA does not directly deplete budgets through direct withdrawals; instead, it optimizes a global fairness constraint that aims to equalize average impression prices across all items.

Implementing a non-auction approach could enhance the system's reliability by mitigating two key pacing limitations: $(1)$ the requirement for additional production services to manage control parameter updates, and $(2)$ the inevitable presence of some pacing imperfections that cannot be fully eliminated.

While our proposed approach shows promise in improving certain metrics in online advertising, we acknowledge that real-world production systems are inherently complex, with numerous products and significant legacy infrastructure. As a result, many companies may be unwilling to abandon auction-based mechanisms. However, OTA can still serve as a valuable offline benchmark tool to analyze the existing fairness-efficiency trade-offs within these systems.

One interesting direction for future research would be to integrate budget-based products with OTA mechanisms, while still utilizing pay-per-click auctions. A possible approach could be to develop a joint optimization framework that combines the strengths of both models, incorporating the benefits of OTA into pay-per-click products.

\bibliographystyle{unsrtnat}
\bibliography{arxivOTA}  

\begin{thebibliography}{22}
\providecommand{\natexlab}[1]{#1}
\providecommand{\url}[1]{\texttt{#1}}
\expandafter\ifx\csname urlstyle\endcsname\relax
  \providecommand{\doi}[1]{doi: #1}\else
  \providecommand{\doi}{doi: \begingroup \urlstyle{rm}\Url}\fi

\bibitem[Chen et~al.(2024)Chen, Nguyen, and Gligorijevic]{Chen2024}
Qinyi Chen, Phuong~Ha Nguyen, and Djordje Gligorijevic.
\newblock Optimization-based budget pacing in ebay sponsored search.
\newblock In \emph{Companion Proceedings of the ACM on Web Conference 2024}, WWW ’24, page 328–337. ACM, May 2024.
\newblock \doi{10.1145/3589335.3648331}.
\newblock URL \url{http://dx.doi.org/10.1145/3589335.3648331}.

\bibitem[Nguyen et~al.(2023)Nguyen, Gligorijevic, Borah, Adalinge, and Bagherjeiran]{Nguyen2023PracticalBP}
Phuong~Ha Nguyen, Djordje Gligorijevic, Arnab Borah, Gajanan Adalinge, and Abraham Bagherjeiran.
\newblock Practical budget pacing algorithms and simulation test bed for ebay marketplace sponsored search.
\newblock In \emph{AdKDD@KDD}, 2023.
\newblock URL \url{https://api.semanticscholar.org/CorpusID:261367675}.

\bibitem[Agarwal et~al.(2014)Agarwal, Ghosh, Wei, and You]{Agarwal2014}
Deepak Agarwal, Souvik Ghosh, Kai Wei, and Siyu You.
\newblock Budget pacing for targeted online advertisements at linkedin.
\newblock In \emph{Proceedings of the 20th ACM SIGKDD international conference on Knowledge discovery and data mining}, KDD ’14. ACM, August 2014.
\newblock \doi{10.1145/2623330.2623366}.
\newblock URL \url{http://dx.doi.org/10.1145/2623330.2623366}.

\bibitem[Karande et~al.(2013)Karande, Mehta, and Srikant]{Karande2013}
Chinmay Karande, Aranyak Mehta, and Ramakrishnan Srikant.
\newblock Optimizing budget constrained spend in search advertising.
\newblock In \emph{Proceedings of the sixth ACM international conference on Web search and data mining}, WSDM 2013. ACM, February 2013.
\newblock \doi{10.1145/2433396.2433483}.
\newblock URL \url{http://dx.doi.org/10.1145/2433396.2433483}.

\bibitem[Stram et~al.(2024)Stram, Abboud, Shtoff, Somekh, Raviv, and Koren]{Stram2024}
Rotem Stram, Rani Abboud, Alex Shtoff, Oren Somekh, Ariel Raviv, and Yair Koren.
\newblock Mystique: A budget pacing system for performance optimization in online advertising.
\newblock In \emph{Companion Proceedings of the ACM on Web Conference 2024}, volume~3 of \emph{WWW ’24}, page 433–442. ACM, May 2024.
\newblock \doi{10.1145/3589335.3648342}.
\newblock URL \url{http://dx.doi.org/10.1145/3589335.3648342}.

\bibitem[Lucier et~al.(2023)Lucier, Pattathil, Slivkins, and Zhang]{Lucier2023AutobiddersWB}
Brendan Lucier, Sarath Pattathil, Aleksandrs Slivkins, and Mengxiao Zhang.
\newblock Autobidders with budget and roi constraints: Efficiency, regret, and pacing dynamics.
\newblock \emph{ArXiv}, abs/2301.13306, 2023.
\newblock URL \url{https://api.semanticscholar.org/CorpusID:256416216}.

\bibitem[Balseiro et~al.(2020)Balseiro, Lu, and Mirrokni]{balseiro2020}
Santiago Balseiro, Haihao Lu, and Vahab Mirrokni.
\newblock Dual mirror descent for online allocation problems.
\newblock In \emph{International Conference on Machine Learning}, pages 613--628. PMLR, 2020.

\bibitem[Yang et~al.(2010)Yang, Vee, Vassilvitskii, Tomlin, Shanmugasundaram, Anastasakos, and Kennedy]{Yang2010InventoryAF}
Jian Yang, Erik Vee, Sergei Vassilvitskii, John~A. Tomlin, Jayavel Shanmugasundaram, Tasos Anastasakos, and Oliver Kennedy.
\newblock Inventory allocation for online graphical display advertising.
\newblock \emph{ArXiv}, abs/1008.3551, 2010.
\newblock URL \url{https://api.semanticscholar.org/CorpusID:612212}.

\bibitem[Dai et~al.(2024)Dai, Lyu, Zhang, Zhao, Zu, Wang, and Zheng]{dai2024}
Liang Dai, Kejie Lyu, Chengcheng Zhang, Guangming Zhao, Zhonglin Zu, Liang Wang, and Bo~Zheng.
\newblock Percentile risk-constrained budget pacing for guaranteed display advertising in online optimization.
\newblock In \emph{Proceedings of the AAAI Conference on Artificial Intelligence}, volume~38, pages 7987--7994, 2024.

\bibitem[Xinying~Chen and Hooker(2023)]{XinyingChen2023}
Violet Xinying~Chen and J.~N. Hooker.
\newblock A guide to formulating fairness in an optimization model.
\newblock \emph{Annals of Operations Research}, 326\penalty0 (1):\penalty0 581–619, April 2023.
\newblock ISSN 1572-9338.
\newblock \doi{10.1007/s10479-023-05264-y}.
\newblock URL \url{http://dx.doi.org/10.1007/s10479-023-05264-y}.

\bibitem[Atkinson(1970)]{Atkinson1970}
Anthony~B Atkinson.
\newblock On the measurement of inequality.
\newblock \emph{Journal of Economic Theory}, 2\penalty0 (3):\penalty0 244–263, September 1970.
\newblock ISSN 0022-0531.
\newblock \doi{10.1016/0022-0531(70)90039-6}.
\newblock URL \url{http://dx.doi.org/10.1016/0022-0531(70)90039-6}.

\bibitem[Bertsimas et~al.(2012)Bertsimas, Farias, and Trichakis]{Bertsimas2012}
Dimitris Bertsimas, Vivek~F. Farias, and Nikolaos Trichakis.
\newblock On the efficiency-fairness trade-off.
\newblock \emph{Management Science}, 58\penalty0 (12):\penalty0 2234–2250, December 2012.
\newblock ISSN 1526-5501.
\newblock \doi{10.1287/mnsc.1120.1549}.
\newblock URL \url{http://dx.doi.org/10.1287/mnsc.1120.1549}.

\bibitem[Bateni et~al.(2022)Bateni, Chen, Ciocan, and Mirrokni]{Bateni2022}
MohammadHossein Bateni, Yiwei Chen, Dragos~Florin Ciocan, and Vahab Mirrokni.
\newblock Fair resource allocation in a volatile marketplace.
\newblock \emph{Operations Research}, 70\penalty0 (1):\penalty0 288--308, 2022.

\bibitem[Balseiro et~al.(2021)Balseiro, Lu, and Mirrokni]{Balseiro2021}
Santiago Balseiro, Haihao Lu, and Vahab Mirrokni.
\newblock Regularized online allocation problems: Fairness and beyond.
\newblock In \emph{International Conference on Machine Learning}, pages 630--639. PMLR, 2021.

\bibitem[Gini(1912)]{gini1912}
C.~Gini.
\newblock \emph{Variabilit{\`a} e mutabilit{\`a}: contributo allo studio delle distribuzioni e delle relazioni statistiche. [Fasc. I.]}.
\newblock Studi economico-giuridici pubblicati per cura della facolt{\`a} di Giurisprudenza della R. Universit{\`a} di Cagliari. Tipogr. di P. Cuppini, 1912.
\newblock URL \url{https://books.google.ru/books?id=fqjaBPMxB9kC}.

\bibitem[Yitzhaki and Schechtman(2012)]{Yitzhaki2012}
Shlomo Yitzhaki and Edna Schechtman.
\newblock \emph{More Than a Dozen Alternative Ways of Spelling Gini}, page 11–31.
\newblock Springer New York, August 2012.
\newblock ISBN 9781461447207.
\newblock \doi{10.1007/978-1-4614-4720-7_2}.
\newblock URL \url{http://dx.doi.org/10.1007/978-1-4614-4720-7_2}.

\bibitem[Lejeune and Turner(2019)]{Lejeune2019}
Miguel~A Lejeune and John Turner.
\newblock Planning online advertising using gini indices.
\newblock \emph{Operations Research}, 67\penalty0 (5):\penalty0 1222--1245, 2019.

\bibitem[Do et~al.(2021)Do, Corbett-Davies, Atif, and Usunier]{do2021two}
Virginie Do, Sam Corbett-Davies, Jamal Atif, and Nicolas Usunier.
\newblock Two-sided fairness in rankings via lorenz dominance.
\newblock \emph{Advances in Neural Information Processing Systems}, 34:\penalty0 8596--8608, 2021.

\bibitem[Do and Usunier(2022)]{do2022optimizing}
Virginie Do and Nicolas Usunier.
\newblock Optimizing generalized gini indices for fairness in rankings.
\newblock In \emph{Proceedings of the 45th International ACM SIGIR Conference on Research and Development in Information Retrieval}, pages 737--747, 2022.

\bibitem[Li et~al.(2024)Li, Rong, Zhang, and Zheng]{Li2024}
Xiaolong Li, Ying Rong, Renyu Zhang, and Huan Zheng.
\newblock Online advertisement allocation under customer choices and algorithmic fairness.
\newblock \emph{Management Science}, 2024.

\bibitem[Chuklin et~al.(2022)Chuklin, Markov, and De~Rijke]{chuklin2022click}
Aleksandr Chuklin, Ilya Markov, and Maarten De~Rijke.
\newblock \emph{Click models for web search}.
\newblock Springer Nature, 2022.

\bibitem[Bubeck et~al.(2015)]{bubeck2015convex}
S{\'e}bastien Bubeck et~al.
\newblock Convex optimization: Algorithms and complexity.
\newblock \emph{Foundations and Trends{\textregistered} in Machine Learning}, 8\penalty0 (3-4):\penalty0 231--357, 2015.

\end{thebibliography}






\end{document}